\documentstyle[epsf]{mn}
\newcommand{\etal}{{\sl et~al.}}

\newcommand{\ro}{{\sl ROSAT}}

\newcommand{\eu}{{\sl EUVE}}

\newcommand{\iue}{{\sl IUE}}
\newcommand{\fu}{{\sl FUSE}}
\newcommand{\fuse}{{\sl FUSE}}
\newcommand{\hs}{{\sl HST}}
\newcommand{\hut}{{\sl HUT}}
\newcommand{\hu}{{\sl HUT}}
\newcommand{\orf}{{\sl ORFEUS}}

\newcommand{\hi}{\mbox{$\rm {H\,{\rm i}}\:$}}

\newcommand{\tlus}{\mbox{$\rm {\rm tlusty}\:$}}
\newcommand{\iraf} {\mbox{$\rm {\rm iraf}\:$}}

\newcommand{\syn}{\mbox{$\rm {\rm synspec}\:$}}
\newcommand{\xsp}{\mbox{$\rm {\rm xspec}\:$}}
\newcommand{\teff}{\mbox{$T_{\rm eff}\:$}}

\title[DA white dwarf temperatures and gravities]
{A comparison of DA white dwarf temperatures and gravities from Lyman
and Balmer line studies}

\author[M.A. Barstow et al.]{
M.A. Barstow$^1$, J.B. Holberg$^2$, I. Hubeny$^3$, 
S.A. Good$^1$,  A.J. Levan$^1$ 
\newauthor and F. Meru$^4$
\\
$^1$ {\it Department of Physics and Astronomy, University of Leicester,
University Road, Leicester LE1 7RH, UK}\\
$^2$ {\it Lunar and Planetary Laboratory, University of Arizona, Tucson, 
AZ 85721, USA}\\
$^3$ {\it Laboratory for Astronomy and Solar Physics, NASA/GSFC, Greenbelt,
Maryland, MD 20711 USA}\\
$^4$ {\it Nottingham High School for Girls, 9 Arboretum Street, Nottingham NG1 4JB, UK} \\
}

\begin{document}

\label{firstpage}

\maketitle

\begin{abstract}

We present measurements of the effective temperatures and surface
gravities for a sample of hot DA  white dwarfs, using the Lyman line
data available from the \hut , \orf  \ and \fu \ far-UV space
missions.  Comparing the results with those from the standard Balmer
line technique, we find that there is a general  good overall
agreement between the two methods. However, significant differences
are found for a  number of stars, but not always of a consistent
nature in that sometimes the Balmer temperature exceeds that derived
from the Lyman lines and in other instances is lower.  We conclude
that, with the latest model  atmosphere calculations, these
discrepancies probably  do not arise from an inadequate theoretical
treatment of the  Lyman lines but rather from systematic effects in
the  observation and data reduction processes, which dominate the
statistical errors in these spectra.  If these
systematic data reduction effects can be adequately controlled, the
Lyman line temperature and gravity measurements 
are consistent with those  obtained from the Balmer
lines when allowance is made for reasonable observational uncertainties. 

\end{abstract}


\section{Introduction}

Two of the most fundamental measurements required to understand the
nature of any star are the determination of its surface gravity and
effective temperature. In the case of the white dwarf stars, where
nuclear burning has long since ceased, temperature is an indication of
their cooling age and, therefore, enables us to map out the
evolutionary sequence.  Early studies of the white dwarf population
relied on photometric measurements (e.g. Koester \etal \ 1979).
However, a major break through in the reliability and accuracy of
these measurements was achieved with the development of a
spectroscopic technique for the H-rich DA white dwarfs, pioneered by
Holberg \etal \ (1986), where the
\iue \ observed \hi \ Lyman $\alpha $ lines of a number of DA white
dwarfs were compared with synthetic model stellar atmospheres. Similar
techniques can be applied to the \hi \ Balmer lines (Holberg \etal \
1985). Indeed, when multiple Balmer lines (usually 3 or 4) are
systematically analysed a unique well-defined determination of both
the temperature and surface gravity can be obtained. Bergeron \etal \
(1992, BSL) were the first to apply this technique to a large sample
of DA white dwarfs. Combining the measurements of \teff \ and $\log g$
with the evolutionary calculations of Wood (1992), BSL obtained
estimates of the mass for each star, confirming the narrowness of the
white dwarf mass distribution and yielding an accurate measurement of
its peak. Subsequent studies of large samples of white dwarfs from
both optical and EUV surveys have firmly established this technique as
the primary way of determining the DA white dwarf temperature scale
and placing these objects in their evolutionary context (see Vennes
\etal \ 1997; Marsh \etal \ 1997; Finley \etal \ 1997a).

The Balmer line technique can only be applied if these features are
visible in the stellar spectrum. However, there are several situations
where this is not the case or where the measurement is compromised in
some way. For example, if a white dwarf resides in a binary system
with a much more luminous main sequence or evolved companion, the
optical spectrum will be dominated by the latter star. A well-known
illustration of this is the DA+K star binary V471 Tauri, which has
been studied extensively (e.g. Barstow \etal \ 1997 and references
therein) and where the Balmer lines are not detectable in the glare of
the K2 dwarf. One major result of the EUV sky surveys has been the
discovery of many more similar systems with companion spectral types
ranging from A to K (e.g. Barstow \etal \ 1994a; Vennes \etal \ 1998;
Burleigh \etal \ 1997).

While the signature of the white dwarf cannot be visually separated
from its companion in these binaries, the UV flux from the degenerate
star dominates that from the late-type star, provided the primary has
a spectral type later than $\approx A2$.  Consequently, most of the
white dwarfs in such systems have been identified from \iue \ spectra.
In principle, the white dwarf temperature and gravity can be
determined from the Lyman~$\alpha $ line profile in each case.
However, a single line is unable to give an unambiguous measurement of
both \teff \ and $\log g$ simultaneously. Additional information, such
as the stellar distance, can constraint the possible values further,
but the distance measurements may not always be sufficiently accurate
or may be based on uncertain knowledge of the primary spectral type.
The importance of access to the Lyman series in an unresolved binary
is illustrated by determination of \teff \ and $\log g$ for several
important white dwarfs including V471 Tauri (Barstow \etal \ 1997) and
HZ43 (Dupuis \etal \ 1998), based on spectra obtained by the \orf \
spectrometer (see below). More recently, Burleigh \etal \ (2001) have
obtained \teff \ and $\log g$ for the white dwarf companion to the A
star $\beta $ Crateris, using \fu .

\begin{figure}
\leavevmode\epsffile{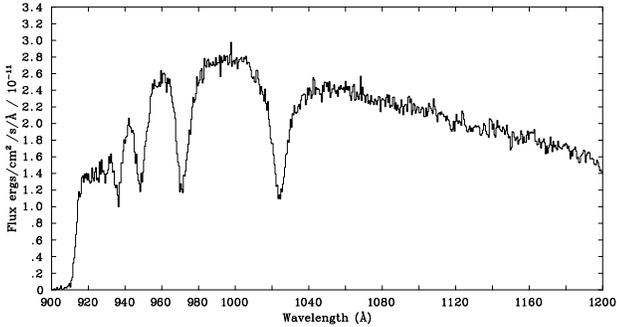}
\caption{Section of the \hut \ spectrum of G191$-$B2B,
covering the wavelength range from 900\AA \ to 1200\AA.
}
\label{hut}
\end{figure}

\begin{figure}
\leavevmode\epsffile{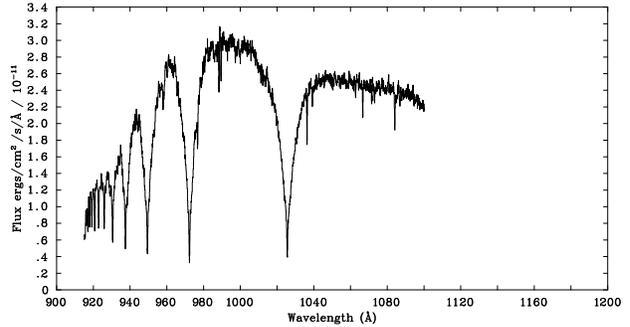}
\caption{\orf \ spectrum of G191$-$B2B.
}
\label{orf}
\end{figure}

In principle, the full Lyman line series could be used in the same
manner as the Balmer lines to determine \teff \ and $\log g$. While
the \iue \ and \hs \ bands do not extend to short enough wavelengths
to encompass more than Lyman $\alpha $, the short duration \hut \ and
\orf \ missions have provided several observations of white dwarfs
down to the Lyman limit with which to examine this idea. The \hut \
telescope was carried into space twice aboard the Space Shuttle, as
part of the {\it Astro 1} and {\it Astro 2} missions. A total of eight isolated
hot DA white dwarfs were observed at a resolution of $\approx 4$\AA \
(fwhm). Launched from the Space Shuttle, the \orf \ missions utilised
a free-flying far-UV spectrometer operating at a factor 10 higher
resolution ($\approx 0.3$\AA ) than \hut . A total of four DAs were
observed during two separate flights.

Several authors have used these far-UV spectra to determine \teff \
and $\log g$. However, as the total number of stars studied on any one
mission has been small, it has been difficult to establish whether or
not the results of Lyman line analyses are generally in agreement with
Balmer line observations of the isolated DA stars. A test of this
nature is crucial if we are to routinely use Lyman line data for this
purpose. A preliminary investigation of this issue was carried out by
Finley \etal \ (1997b), using the {\it Astro 2} \hut \ data. Their
major conclusion was that the Lyman lines were significantly weaker
than the standard Stark broadening theory of Vidal \etal \ (1973).
Conflicting results have been obtained with the \orf \ data. For
example, Barstow \etal \ (1998), utilizing state-of-the-art non-LTE
models, found that the Lyman and Balmer \teff \ determinations were in
good agreement for the heavy element rich star G191$-$B2B but not for
the similar object REJ0457$-$281. Dupuis \etal \ (1998) also achieved
good agreement between the results of their Lyman line analysis of the
pure H atmosphere white dwarf HZ43 and other published values of \teff
\ from a variety of published sources including \hut , \eu \ besides
ground-based Balmer line observations. However, it must be remembered
that these measurements reported in the literature were derived using
several different model codes, spanning several generations of these
programmes. Hence, such comparisons lack the uniformity and
self-consistency required to evaluate the efficacy of the Lyman lines
as reliable \teff \ and $\log g$ indicators. The launch of the Far
Ultraviolet Spectroscopic Explorer (\fu ) on 1999 June 24 has provided
long duration access to the Lyman series region of the electromagnetic
spectrum for the first time since Copernicus, in the early 1970s.
However, Copernicus was not sufficiently sensitive to observe any
white dwarfs, except Sirius B. With a spectral resolution superior to
both \hut \ and \orf , and comparable effective area, \fu \ promises
to produce Lyman series data for many white dwarfs. This paper
presents a critical analysis of the use of the H Lyman series
absorption lines to determine \teff \ and $\log g$. We reanalyze the
archival \hut \ and \orf \ spectra in conjunction with new, improved
non-LTE synthetic spectra, comparing the results with our archive of
Balmer line data, using the same models.  We also consider the
possible systematic observation and data reduction effects that might
affect either or both Balmer and Lyman procedures.

\section{Observations}

\subsection{Lyman line spectra}

We have obtained far-UV spectra covering the Lyman series from the
$\beta $ line through to the series limit from three telescopes, \hut,
\orf \ and \fu . Table~\ref{fuv} summarises all the observations,
which are discussed briefly below. All the spectra were obtained from
the Multi-mission archive (MAST, http://archive.stsci.edu/mast.html)
hosted by the Space Telescope Science Institute, or directly from the
investigators of the respective instruments.

\begin{table}
\caption{Far-UV Lyman series spectra 
used in this study, obtained from \hut , 
\orf \ and \fu \ missions.}
\begin{tabular}{lll}
Star & Instrument & Data set name/number \\
GD50 & \hut 2 & A19001, A19002\\
GD394 & \hut 2 & A09201, A09202, A13801, A21101\\
HZ43         & \hut & A32901, A32902 \\
        & \orf 2  & 12, 3270659-3362240 \\
REJ0512 & \hut 2 & A12701, N12701 \\
G191$-$B2B                     & \hut 2& N11401 \\
                    & \orf 1& 4, g191b2b$\_$1-4\\
GD153            & \fu & M1010401 \\
GD71 & \hut 2 & N11201 \\
   & \fu & M1010301, M1010302 \\
         
REJ0457 & \orf 1& 4, mct0455$\_$1-4\\
PG1342+444 & \fu & A0340402 \\
REJ0558 & \fu & A0340701 \\
REJ1738  & \fu & A0340301 \\
\label{fuv}
\end{tabular}
\end{table}

\begin{figure}
\leavevmode\epsffile{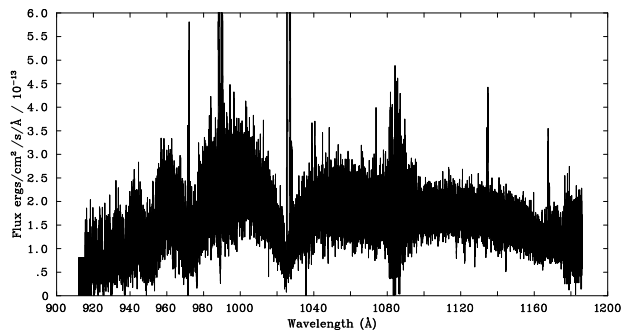}
\caption{\fu \ spectrum of PG1342+444, a) combined
from the individual grating spectra, before applying our processing
pipeline.}
\label{fuse}
\end{figure}

\begin{figure}
\leavevmode\epsffile{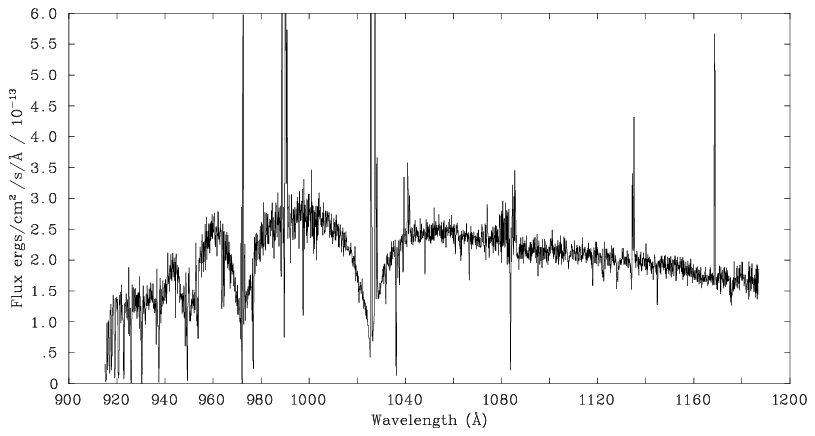}
\contcaption{b)  After processing, as described in the text and
resampled with a  0.1\AA \ bin size to improve the s/n.}
\end{figure}

\subsubsection{\hut \ observations}

The \hut \ instrument was flown on two space shuttle missions, {\it
Astro 1} and {\it Astro 2}, in 1990 December and 1995 March
respectively. The payload consisted of a 90cm f/2 telescope with a
Rowland circle spectrograph at the prime focus, covering the
820-1840\AA \ wavelength range, in first order. The spectral
resolution ($\lambda/\Delta \lambda =500$) was dependent upon the
instrument temperature and pointing stability. The final calibration
is described by Kruk \etal \ (1997).The second flight incorporated
significant improvements in sensitivity and stability. The spectral
resolution was wavelength dependent in this flight, varying from 2-4.5
\AA \ (fwhm). Both temporal and wavelength dependent sensitivity
changes were experienced during the mission, which were monitored and
characterized by multiple observations of three white dwarfs. Kruk
\etal \ (1999) discuss the final {\it Astro 2} \hu \ calibration. An
example \hut \ spectrum, of G191-B2B, is shown in figure~\ref{hut}
While eight stars were observed, in REJ1738 strong H$_2$ absorption prevents
sensible analysis of the H Lyman lines
(the increased resolution allows the H$_2$ lines
to be removed in the \fu \ spectrum). In addition, Wolf 1346 is too
cool to show a full Lyman line series and is not suitable for this
study. Hence, we only made use of observations of six stars.

\begin{figure}
\leavevmode\epsffile{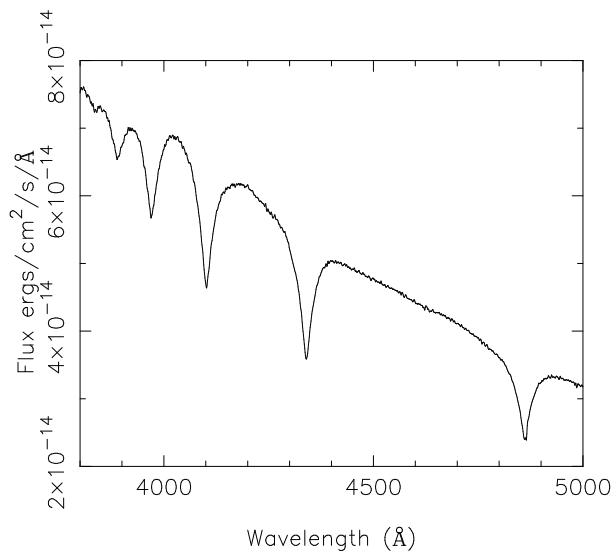}
\caption{Ground-based optical spectrum of GD71.}
\label{gb}
\end{figure}

\subsubsection{\orf \ observations}

Higher resolution far-UV spectra were obtained with the \orf \
telescope (Hurwitz \& \ Bowyer 1991) on the free-flying {\it
Astro-SPAS} platform, deployed from the shuttle. \orf \ was flown on
two occasions, a 5 day mission in 1993 September and a 14 day mission
in 1996 November. On the first flight, the Berkeley spectrometer
located at the focus of the 1-m primary, covered the spectral range
360-1176\AA \ at a resolution of $\lambda/\Delta \lambda=5000$
(Hurwitz \& \ Bowyer 1995; Hurwitz \etal \ 1998). For the second
mission, the far-UV wavelength coverage was 900-1200\AA \ and the
spectral resolving power 3300 (Hurwitz \etal \ 1998).

All data obtained during both \orf \ missions are now in the public
domain. The raw spectra suffer from contamination by a scattered light
component, which is a combination of EUV flux from the second spectral
order and direct scatter from the grating (Hurwitz, private
communication). When added to the stellar spectrum, the apparent level
of the continuum flux will be higher than the true one. In
determinations of \teff \  and $\log g$, the line profiles, whether
Balmer or Lyman, are measured from the continuum level. Therefore, the
scattered light component must be accounted for in any analysis to
avoid obtaining erroneously high temperatures from apparently weaker
lines. Immediately shortward of the Lyman limit, the stellar spectrum
makes no contribution to the net flux, as a result of absorption by
interstellar neutral hydrogen. Therefore, the level of the scattered
light component was estimated from the observed flux in the
850--900\AA \ range. The \orf \ spectrum of G191$-$B2B is shown in
figure~\ref{orf}, for comparison with the \hut \ spectrum
(figure~\ref{hut}). While four stars were observed, one of these, the
white dwarf in V471 Tauri is in a close binary and no Balmer series data 
can be obtained due to the overwhelming brightness of the K star
companion.

\subsubsection{\fu \ spectra}

The \fu \ mission was placed in low Earth orbit on 1999 June 24. After
several months of in-orbit checkout and calibration activities,
science observations began during 1999 December. An overview of the
\fu \ mission has been given by Moos \etal \ (2000) and the
spectrograph is described in detail by Green \etal \ (1994). Further
useful information is included in the \fu \ Observer's Guide (Oegerle
\etal \ 1998) which can be found with other technical documentation on
the \fu
\ website
(http://fuse.pha.jhu.edu).

Only a few scientific papers have, as yet, been published
incorporating \fu \ data. Hence, it is appropriate to give a brief
description of the spectrometer and its current status in the context
of the reduction and analysis of the data presented here. The far-UV
spectrometer is based on the Rowland circle design and comprises four
separate optical paths (or channels). Each channel consists of a
mirror, a focal plane assembly (including the spectrograph apertures),
a diffraction grating and a section of one of two detectors. To
maximise the throughput of the instrument, the channels must be
co-aligned so that light from a single target properly illuminates all
four channels. Two of the mirrors and two of the gratings are coated
with SiC to provide sensitivity at wavelengths below $\approx 1020$\AA
, while the other two mirror/grating pairs are coated with Al and a
LiF overcoat. This latter combination yields about twice the
reflectivity of SiC at wavelengths above 1050\AA , but has little
reflectivity below 1020\AA. The overall wavelength coverage runs from
905\AA \ to 1187\AA.

Spectra from the four channels are recorded on two microchannel plate
detectors, with a SiC spectrum and LiF spectrum on each. The
individual detectors are divided into two functionally independent
segments (A and B), separated by a small gap. Consequently, there are
eight detector segment/spectrometer channel combinations to be dealt
with in reducing the \fuse \ data. The nominal wavelength ranges of
these are listed in table~\ref{lrange}.

\begin{table}
\caption{Nominal wavelength ranges (\AA ) for the \fuse \ detector segments}
\begin{tabular}{lll}
Channel & Segment A & Segment B \\
SiC 1 & 1091.1--1003.9 & 992.6--905.0\\
LiF 1 & 987.1--1082.2 & 1094.3--1187.7\\
SiC 2 & 916.8--1006.4 & 1016.3--1105.0\\
LiF 2 & 1181.7--1085.6& 1074.6--978.1\\
\end{tabular}
\label{lrange}
\end{table}

Several problems have been reported during in-orbit operations of the
\fu \ satellite which need to be taken into account in the data
reduction process. Maintaining the co-alignment of the individual
channels has been difficult, probably due to thermal effects.
Sometimes, during an observation, a target may completely miss the
aperture of one or more channels, while being well centred in the
others. In addition, even if the channels are well-aligned at the
beginning of an observation, the target may subsequently drift out of
any of the apertures. To minimize this problem, most observations have
been conducted using the largest aperture available (LWRS, $30\times
30$ arcsec), limiting the spectral resolution for point sources to
between 10000 and 20000 ($\approx 0.05$ to 0.1\AA) 
compared to the 24000-30000 expected for the
$1.25\times 20$ arcsec HIRS aperture.

Most of the \fu \ spectra included in this work were obtained in
time-tag mode (where the arrival time of each detected photon is
recorded and the image subsequently reconstructed from the positional
data also included in the data stream) in several separate exposures
through the LWRS aperture (table~\ref{fuv}). However, two observations
(GD71 and GD153) were obtained from the MAST archive as part of the
early release of data for calibration purposes. All these data were
processed with the CALFUSE pipeline version 1.6.6. Subsequently, a new
wavelength calibration has been made available and which is applied to
all data processed recently (pipeline version 1.7). We have replaced
the original wavelength calibration files with the revised versions
according to the procedure described on the \fuse \ data analysis web
pages (fuse.pha.jhu.edu/analysis/wavelength$\_$062200.html), before
combining the individual spectra.

Initially, for each star, we considered the separate exposures for a
single channel/detector segment. Since, the signal-to-noise of these
is relatively poor and the wavelength binning ($\approx 0.006$\AA)
over samples the true resolution by a factor of 2-3, all the spectra
were re-binned to 0.02\AA \ pixels for examination. Fortuitously,
several prominent interstellar absorption lines are detected in each
spectrum showing that, for a given channel and detector side, the
wavelength scales of each exposure are well aligned. Consequently, it
is a straightforward process to co-add the individual exposures, to
produce a single spectrum.  We used the \iraf \ script FUSECOMBINE
(see fuse.pha.jhu.edu/analysis/IRAF$\_$scripts.html) to co-add the
multiple exposures, which weights the individual spectra by their
exposure time. This whole procedure was repeated for all eight
channel/detector side combinations for each star.

Since the complete \fu\ wavelength range is covered in 3 $\approx
100$\AA \ bands, a number of the individual detector segments overlap
almost completely in wavelength. For example, the SiC 1A, SiC 2B, LiF
2A and LiF 2B all cover the range $\approx 1000-1100$\AA , while the
SiC 2A/2B ($\approx 900- 1000$\AA ) and LiF 1A/1B ($\approx
1100-1200$\AA ) match in wavelength. Consequently, to achieve the
optimum signal-to-noise for analysis of the data, it is desirable to
combine all these individual segment spectra. We have written a small
Fortran programme to do this, which is able to take into account the
differing wavelength ranges and spectral bin sizes of each. First, all
the spectra are re-sampled onto a common wavelength scale and then
re-binned into 0.02\AA \ steps to reduce the level of over sampling
and avoid any fringing effects that might arise from the first part of
the procedure. The re-sampled/re-binned spectra are then co-added,
weighting individual data points by the statistical variance, averaged
over a 10\AA \ interval, to take into account the differences in
effective area of each segment and any differences in exposure time
that may have arisen from rejection of bad data segments. We find,
through visual inspection of the observed spectra, that the
statistical noise tends to increase towards the edges of the
wavelength range. In cases where the signal-to-noise is particularly
poor in these regions, we have trimmed the spectra to remove these
data points prior to co-addition. 

As an example, the \fu \ spectrum
of PG1342+444 is shown in figure~\ref{fuse}a, combined from the
individual grating spectra but without any of the processing described
above. The strong high flux spikes are geocoronal emission lines, while
poor s/n regions of the spectrum (usually covered by only one grating)
are revealed by increased scatter in the data points (e.g. near
1080\AA). A few of the stronger heavy element and interstellar
absorption lines are also visible. The broad dip in the 1160-1180\AA \ 
region of the spectrum is an artefact known as the ``worm", which is
a local 10-20\% \ reduction in the effective area and produces a
corresponding loss of flux. Figure~\ref{fuse}b shows the same
spectrum processed according to our prescription and resampled with a
bin size of 0.1\AA \ to improve the s/n for our analysis.

\subsection{Balmer line spectra}.

The majority of the optical data we use here was obtained as part of a
spectroscopic follow-up programme following the \ro \ X-ray and EUV
sky survey. Observations were undertaken in both Northern and Southern
hemispheres. Southern hemisphere data were obtained with the 1.9-m
Radcliffe reflector of the South African Astronomical Observatory
(SAAO), while stars in the Northern Hemisphere were observed with the
Steward Observatory 2.3-m telescope on Kitt Peak. Full details of
these observations have already been published by Marsh \etal \
(1997). The main difference between the Southern and Northern
Hemisphere data is their spectral resolution, $\approx 3$\AA \ (FWHM)
and $\approx 8$\AA \ (FWHM) respectively. Some of our original
Northern Hemisphere spectra did not cover the complete Balmer line
series, excluding H$\beta $, due to the limited size of the CCD chip
available at the time. Hence, in those cases, we have replaced the archival data
with more recent observations (GD394 and GD71) made using the same
instrument but with a larger chip (see e.g. figure~\ref{gb}). For some
stars (GD50, HZ43, G191$-$B2B, and GD153), while archival (``old") data  
are of good quality and cover all the Balmer line series the ``new"
observations have longer exposures and, consequently, better s/n. 
We analyse all these data, as they provide a useful test of consistency
between repeated observations of the same star, using basically the same
instrument configuration but on different nights.

%
%
\begin{figure}
\leavevmode\epsffile{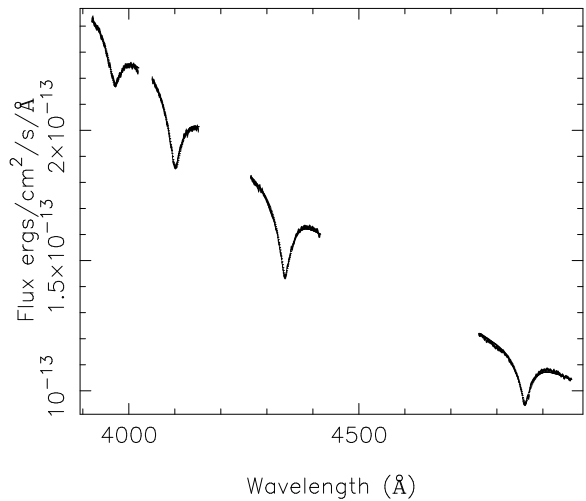}
\caption{Fit to the ground-based optical H$\beta $ to H$\epsilon $ lines of
G191$-$B2B (\teff $=51510\pm 880$K, $\log g=7.53\pm 0.09$).}
\label{lines}
\end{figure}


\begin{figure}
\leavevmode\epsffile{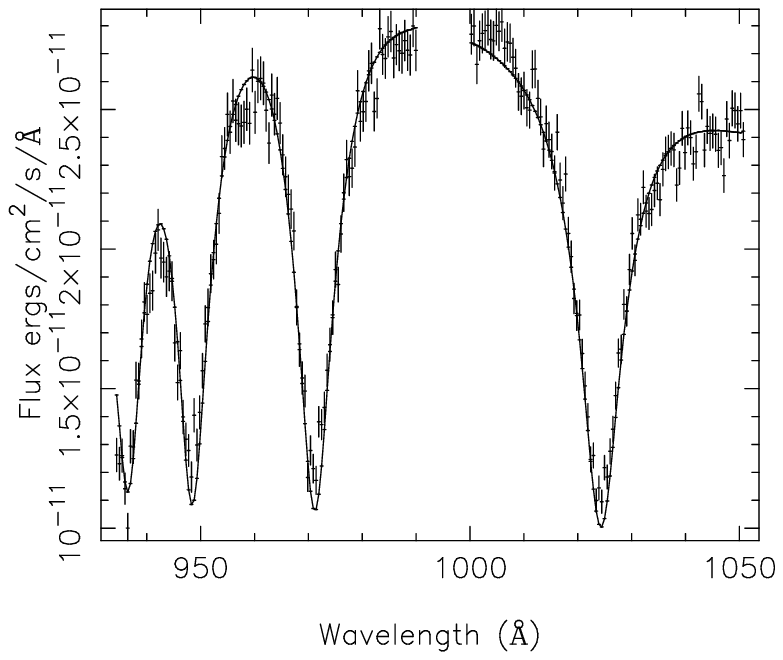}
\contcaption{ b) Fit to the \hut \  L$\beta $
to L$\epsilon $ lines of G191$-$B2B (\teff $=52930\pm 360$K,
$\log g=7.16\pm 0.20$). 
Although the line cores are not
included in the analysis, they are shown
here for illustrative purposes.}
\end{figure}

\begin{figure}
\leavevmode\epsffile{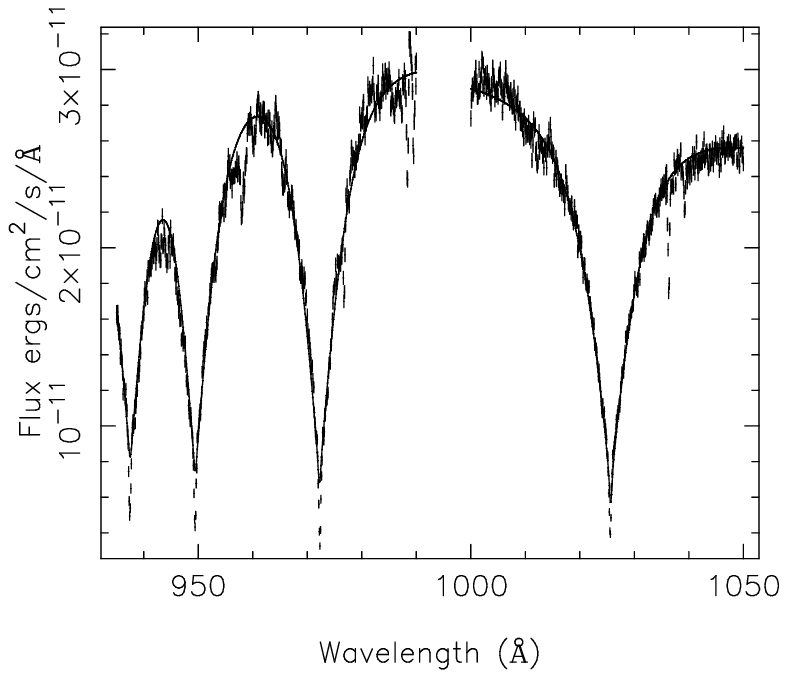}
\contcaption{ c) Fit to the \orf \ L$\beta $ 
to L$\epsilon $ lines of G191$-$B2B
(\teff $=53180\pm 520$K, $\log g=7.43\pm0.04$).
Although the line cores are not included in the analysis, they are shown
here for illustrative purposes.
 }
\end{figure}

\section{Model Atmosphere Calculations}

One potential flaw in comparing the results from published Lyman and
Balmer line analyses is that different authors may have utilized
different stellar atmosphere codes, even spanning successive
generations of these codes. To ensure that our work is at least
internally consistent we re-examine all the Balmer line data with the
latest version of the stellar atmosphere programme \tlus \ and its
associated spectral synthesis package \syn .

We have calculated completely new grids of model stellar atmospheres
using the non-LTE code \tlus (Hubeny \& \ Lanz 1995). These are based
on work reported by Lanz \etal \ (1996) and Barstow \etal \ (1998,
1999). Two separate sets of calculations were performed: pure H
models, for those stars without significant abundances of heavy
elements, and models with a homogeneous mixture of heavier elements,
including C, N, O, Si, Fe and Ni, for the others. The upper limit to
the temperature of the pure H models was 70000K, to span the range
within which DAs with pure H atmospheres are found, while the heavy
element calculations were extended to 120000K, to deal with hotter
DA stars, although we note that in this particular study
the hottest star we consider is REJ1738, with \teff $\approx 70000$K. 
To take account of the higher
element ionization stages that are likely to be encountered in the
latter objects, we have added new ions of OVI, FeVII/VIII and
NiVII/VIII to the model atoms as well as extending the data for
important ions such as CIV to include more energy levels. As before,
all the calculations were performed in non-LTE with full
line-blanketing, including Stark broadening of all the CNO lines.

The stars included in this study are divided into two distinct groups.
Those with pure H envelopes, for which we used the pure H model
calculations, and those with significant quantities of heavy elements.
For the latter group, we
fixed the abundances of the heavy elements at the values
determined from our earlier homogeneous analysis of G191$-$B2B
(He/H=$1.0\times 10^{-5}$, C/H=$4.0\times 10^{- 7}$, N/H=$1.6\times
10^{-7}$, O/H=$9.6\times 10^{-7}$, Si/H=$3.0\times 10^{-7}$,
Fe/H=$1.0\times 10^{-5}$, Ni/H=$5.0\times 10^{-7}$), but taking into
account that the CIV lines near 1550\AA \ have subsequently been
resolved into multiple components by STIS (Bruhweiler \etal \ 2000;
Bannister \etal \ 2001). While not all stars have exactly the same
heavy element abundances, our recent work has shown that the differences
are not very large (see Barstow \etal \ 2001) and, at this level, will
not have a significant effect on the Balmer/Lyman line measurements
(see e.g Barstow \etal \ 1998).

\begin{figure*}
\leavevmode\epsffile{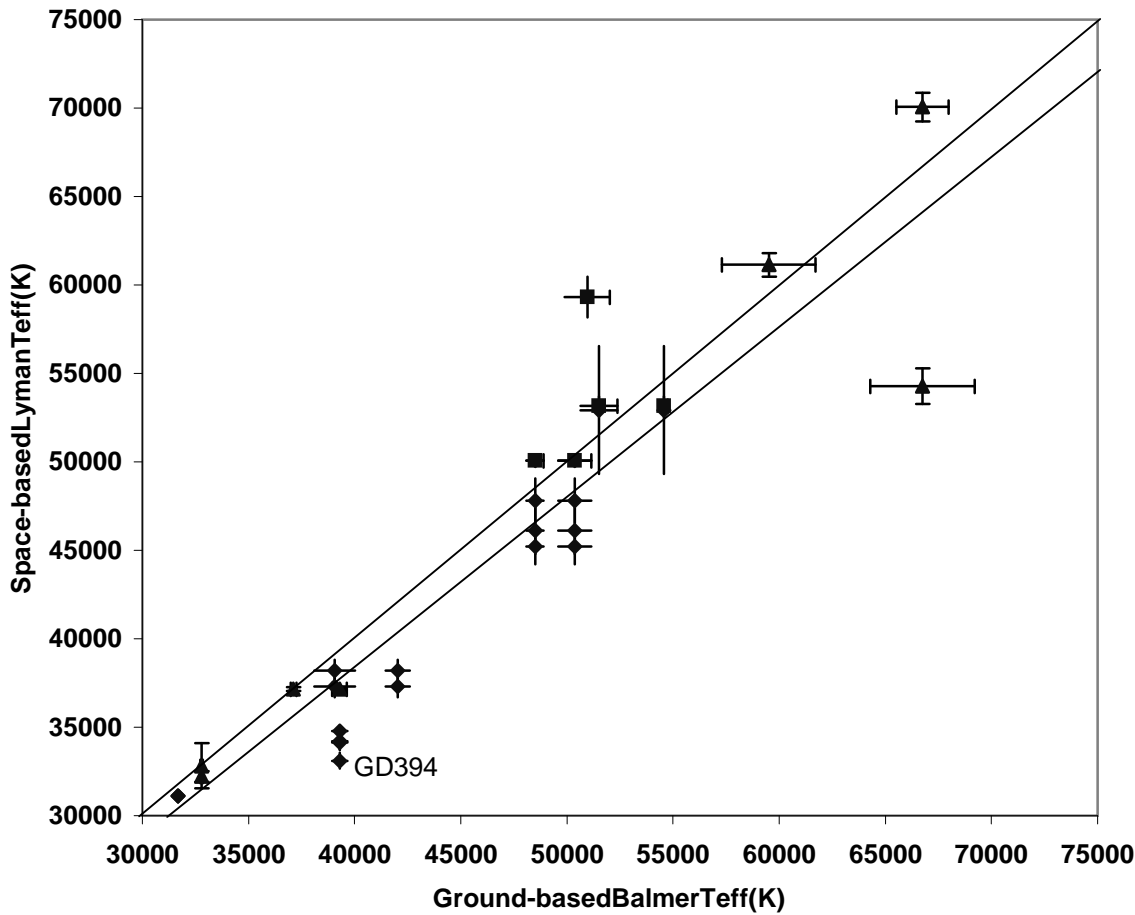}
\caption{Scatter
plot of the values of \teff \ measured using the Balmer and Lyman
series lines. Diamonds are \hut \ data, the squares are the \orf \ data
and the triangles the \fu \ data. The error bars displayed
correspond just to the statistical $1\sigma $ uncertainties. 
The upper solid line corresponds to equal Balmer and
Lyman temperatures, while the lower line is the least squares straight
line fit to the data, corresponding to y=0.964x.}
\label{all}
\end{figure*}

In the spectrum synthesis code \syn (Hubeny \etal \ (1994), we have
replaced the hydrogen Stark line broadening tables of Sch\"oning \& \
Butler (private communication) by the more extended tables of Lemke
(1997). The latter allow a more accurate interpolation of the electron
density for high density environments, such as the atmospheres of
white dwarfs. The spectra produced by the \tlus /\syn \ codes were
recently extensively tested against the results of Koester's codes
(Hubeny \& \ Koester in preparation). The differences in the predicted
spectra for \teff =60000K and $\log g=8$ were found to be below 0.5\%
\ in the whole UV and optical range. Furthermore, we have found that
the inaccuracy in the interpolations of the Sch\"oning and Butler
tables, together with some fine details of our treatment of the level
dissolution, were the primary reason for the disagreement between the
spectroscopically deduced \teff \ using \tlus and Koester models
obtained by Barstow \etal \ (1998). These changes largely resolve the
differences between codes noted by Bohlin (2000).

\section{Determination of Temperature and Gravity}

The technique for determining \teff \ and $\log g$, by comparing the
line profiles with the predictions of synthetic spectra is well
established (see Holberg \etal \ 1986; Bergeron \etal \ 1992 and many
subsequent authors). We have described our own Balmer line analysis
technique in several earlier papers (e.g. Barstow et al. 1994b), but
as the results presented in this paper rely heavily on it we repeat
the details here.  The same technique can also be applied to analysis
of the Lyman lines, as has already been demonstrated (e.g. Barstow
\etal \ 1997; Barstow \etal \ 1998). However, we have modified our
earlier approach, as outlined below.

Both sets of \teff \ and $\log g$ measurements were performed using
the program \xsp \ (Shafer \etal \ 1991), which adopts a $\chi^{2}$
minimization technique to determine the model spectrum which gives the
best agreement with the data. For the Balmer lines, the four strongest
($\beta , \gamma , \delta ,
\epsilon $) are simultaneously included in the fit and an independent
normalisation constant applied to each, ensuring that the result was
independent of the local slope of the continuum and reducing the
effect of any systematic errors in the flux calibration of the
spectrum. Individual lines maintain their local continuum slope but
are decoupled from the overall energy distribution of the entire
Balmer line region.

In the case of the Lyman lines, the analysis needs to be handled in a
slightly different way. In general, at given
values of \teff \ and $\log g$, the Lyman lines are stronger, and,
apart from the $\alpha $ and $\beta $ lines, overlap substantially. To
deal with the Lyman data for each instrument, we separated out the
$\beta $ (1000- 1050\AA ) line and incorporated the remaining lines
($\gamma $ through $\epsilon $ inclusive) into a single file.  To take
account of any low frequency systematic effects in the flux
calibration we applied individual normalization constants to each of
the two sections of data.

Since the Lyman data are, by definition, obtained
from space-based platforms, there are no systematic errors arising
from an atmospheric extinction correction. Furthermore, the flux
calibration is usually obtained from a detailed off-line calibration
of the instrument, applied as part of a standard pipeline, rather than
direct comparison of the observed spectrum with that of a selected
standard star. However, the instrument calibration still refers to
observations of well-studied stars, such as white dwarfs, some
systematic uncertainties will still apply, but will be different to
those arising from the ground-based techniques
Although, in general, no extinction correction needs to be applied to
the Lyman series data, there are two important effects that need to be
accounted for in any analysis. First, the cores of the Lyman
absorption lines can be contaminated by geocoronal emission
components. Second, interstellar absorption can artificially deepen
the core of the stellar Lyman absorption lines. These two effects
compete with each other and may occasionally conspire to cancel each
other out, but usually they must be removed from the data in an
appropriate way.

\begin{figure*}
\leavevmode\epsffile{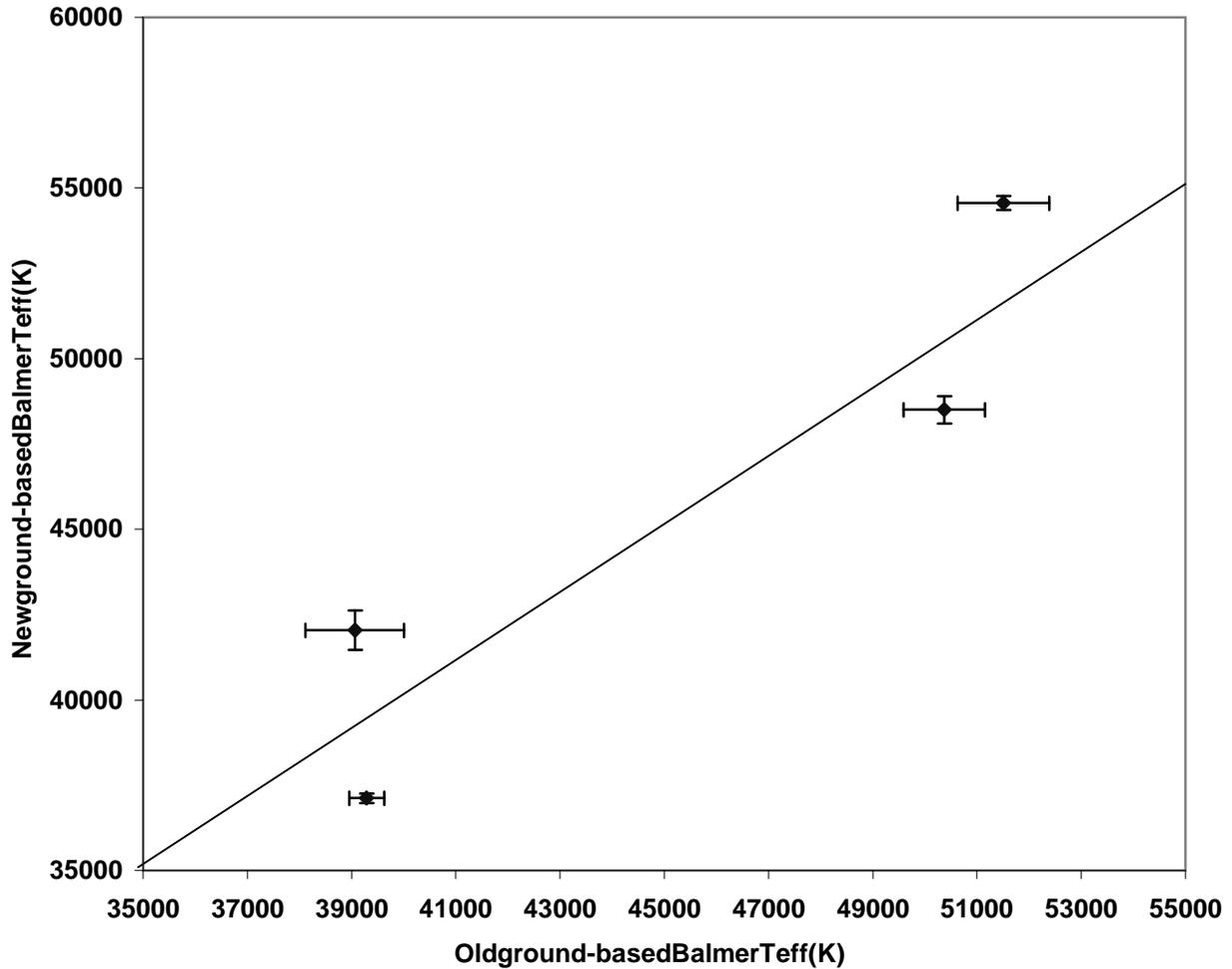}
\caption{Scatter
plot of the values of \teff \ measured using the ``old" (x-axis) and
``new" (y-axis) Balmer series lines. The error bars displayed
correspond to statistical $1\sigma $ uncertainties alone. The light
solid line corresponds to equal Balmer temperatures.}
\label{gbgb}
\end{figure*}

The observed strengths of individual geocoronal lines depend on a
number of factors, including the strength of the Solar line being
scattered, the density of scattering atoms along the line of site
(here \hi \ atoms) and the optical depth of the scattering
environment. In general however, the strongest line is always Lyman
$\alpha $ followed by Lyman $\beta $. In most of 
the data considered here,
the Lyman $\alpha $ line is not observed and is only available
in the \hut \  data. Hence, it is not used for
consistency of analysis and the Lyman $\beta $ line is by far the
strongest component we need to consider 
(e.g. Figure~\ref{fuse}). From a signal-to-noise
point of view the relative strength of the emission when compared to
the stellar continuum is also affected by the viewing geometry, the
instrument design and the brightness of the target star. The
geocoronal radiation has a natural spectral width caused by Doppler
effects, which is then further broadened in the instrumental aperture
by the diffuse nature of the source and the spectral resolution of the
instrument. As a result of possible differences in the relative
velocity of the stellar and geocoronal sources, the emission lines are
not necessarily aligned with the stellar absorption cores.


To make sensible estimates of the uncertainty in the fitted
parameters, the value of the reduced $\chi^2$ should be less than
$\approx 2$. This is the case for the fits to all the
spectra included in this analysis, which are, therefore, formally
good fits. Errors on \teff \ and $\log g$ can then determined by
allowing the model parameters to vary until the ${\delta}{\chi}^{2}$
reached the value corresponding to the $1\sigma$ level for 2 degrees
of freedom (2.3). It should be noted that these estimates only include
statistical uncertainties and do not take into account any possible
systematic effects related to the data acquisition and reduction
processes.

Figure~\ref{lines} shows examples of the fits to the Balmer (5a)
and Lyman (5b, 5c) lines of G191$-$B2B respectively. G191$-$B2B is one
of only two stars (the other is HZ43), where spectra exist for
ground-based, \hut \ and \orf , providing a useful cross-check
between all these instruments. This is particularly important as,
prior to the launch of \fu , the majority of the Lyman line DA data
had been obtained by \hu .

\begin{table*}
\caption{
Best-fit effective temperature and surface gravity for each
observation of each DA white dwarf in the sample. Multiple
values given each column refer to independent observations
using the same instrument.}
\begin{tabular}{lllllllll}
STAR	& GROUND &&\hut &&\orf (o)/\fu (f)\\	
        & BASED \\
	&\teff (K) (err) & $\log g$ (err)  & \teff (K) (err) 
& $\log g$ (err) & \teff  (K) (err )& $\log g$ (err) \\
					
GD50	&39060(950)	&9.30(0.12)	 &37300(600)&8.81(0.19)	\\ 	 		 		 
	&42040(570)	&9.15(0.05)	 &38200(600)&8.82(0.18)	\\ 		 		 		 
\\
GD394	&39290(360)	&7.89(0.05)	 &34790(300)&7.96(0.10)	\\ 		 		 		 
	 &&&33100(450)	&7.63(0.15)	 \\		 		 		 
	 &&&34200(500)	&8.01(0.10)	 \\		 		 		 
	 &&&34150(400)	&8.00(0.10)       \\ 				 		 
\\
HZ43	&50370(780)	&7.85(0.07)&46100(1300)&7.65(0.09)	&(o)50080(290)&8.14(0.05)	\\  
	&48500(400)	&8.05(0.02)&47800(1250)&7.84(0.12)\\	
        &&&45200(1000)	&7.68(0.13)	 \\ 
\\        
RE0512&	31670(140)	&7.20(0.04)	 	&	 		31100(250)	&7.21(0.10)
	\\ 		 		 		 
	& &&		 		 		 		31100(350)	&7.19(0.14)	 \\		 
\\
G191B2B	&51510(880)	&7.53(0.09)&52930(3600)	&7.16(0.20)	&(o)53180(520)	
&7.43(0.04)	 \\		 
	&54560(200)	&7.60(0.02)    	  	\\	 
\\
GD153	&39290(340)	&7.77(0.05)	 		 	
	 &&&		 		(f)37150(100)	&8.1(0.03)\\
	&37120(140)	&8.02(0.02)		 \\		 	
\\
GD71	&32780(65)	&7.83(0.02)		 	 	
	& 31970(150)& 7.72(0.07)& 		 		(f)32820(1280)	&8.85(0.28)\\
	& &&& &		(f)32220(310)	&8.49(0.07)\\
\\
REJ0457	&50960(1070)	&7.93(0.08)	 &&&		 		 		 
		(o)59300(1150)	&7.57(0.08)	 \\		 	 
\\
PG1342	&66750(2450)	&7.93(0.11)	 		 		 		 	
	 &&&		 		(f)54286(1000)	&7.82(0.07)\\
\\
REJ0558	&59510(2210)	&7.70(0.14)	 		 		 		 	
	 &&&		 		(f)61140(660)	&7.61(0.05)\\
\\
REJ1738	&66760(1230)	&7.77(0.10)	 		 		 		 	
	 &&&		 		(f)70060(800)	&8.00(0.01)\\

\label{temps}
\end{tabular}
\end{table*}


Table~\ref{temps} summarises the effective temperatures and gravities
obtained from the spectral fits for all data sets for each star,
including the $1~\sigma $ statistical uncertainties on these values.
Where more than one ground-based balmer line measurement is reported,
the upper value is obtained with the ``old" archival data and the lower
is from the ``new", higher s/n observation. For GD394 and GD71, the
reported measurements are only from the ``new" spectra as the archival
obervations only span 2-3 lines of the Balmer series and are not used
here.

The \teff \ and $\log g$ measurements obtained from the various
observations are a somewhat heterogeneous group. While we have a
complete set of ground-based optical spectra for all the stars, and in
some cases multiple ground-based observations of an individual star,
the far-UV observations are incomplete, with different instruments
covering a different group. At this stage, for no star is there a
complete set of observations with data from all space-based
instruments. In a number of cases, there are multiple observations of
a star made with a single telescope. Rather than combine these spectra
into a single average, where the signal-to-noise is adequate, we have
chosen to analyse these individually to see if any systematic
differences may occur between separate observations of the same star
with the same instrument. We discuss the comparison of individual
groups of data below.

\section{Discussion}

\subsection{Effective temperatures}

In figure~\ref{all}, we show a scatter plot of the values of \teff \
measured by Balmer line (x-axis) and  Lyman line (y-axis)
techniques. All
combinations of ground and far-UV observations are shown, to
illustrate  the full range of possible systematic differences. Also
shown is the straight line corresponding to equal  Balmer and Lyman
temperature together with the best fit straight line corresponding to
y=0.964x. In  general, there is good agreement between the two
measurement techniques, but there also exist some  clear
anomalies. For example, the Lyman line values of \teff \ for GD394 are
consistent with each other,  but systematically lower than the Balmer
line value (\teff \ balmer = 39290) by $\approx 5000$K.  Similarly,
the value of \teff \ measured from the Balmer lines of PG1342+444 is
12000K higher than the  corresponding Lyman line measurement. However,
such differences are not all in the same  direction. The Lyman
line temperature of REJ0457 is some 8000K higher than the Balmer
temperature. In general,  the Lyman
\teff \  measurements are about 4\% \ below the Balmer line values. At
50000K, this corresponds to a  difference of 2000K, which is of a
similar magnitude to many observational errors.

\begin{figure*}
\leavevmode\epsffile{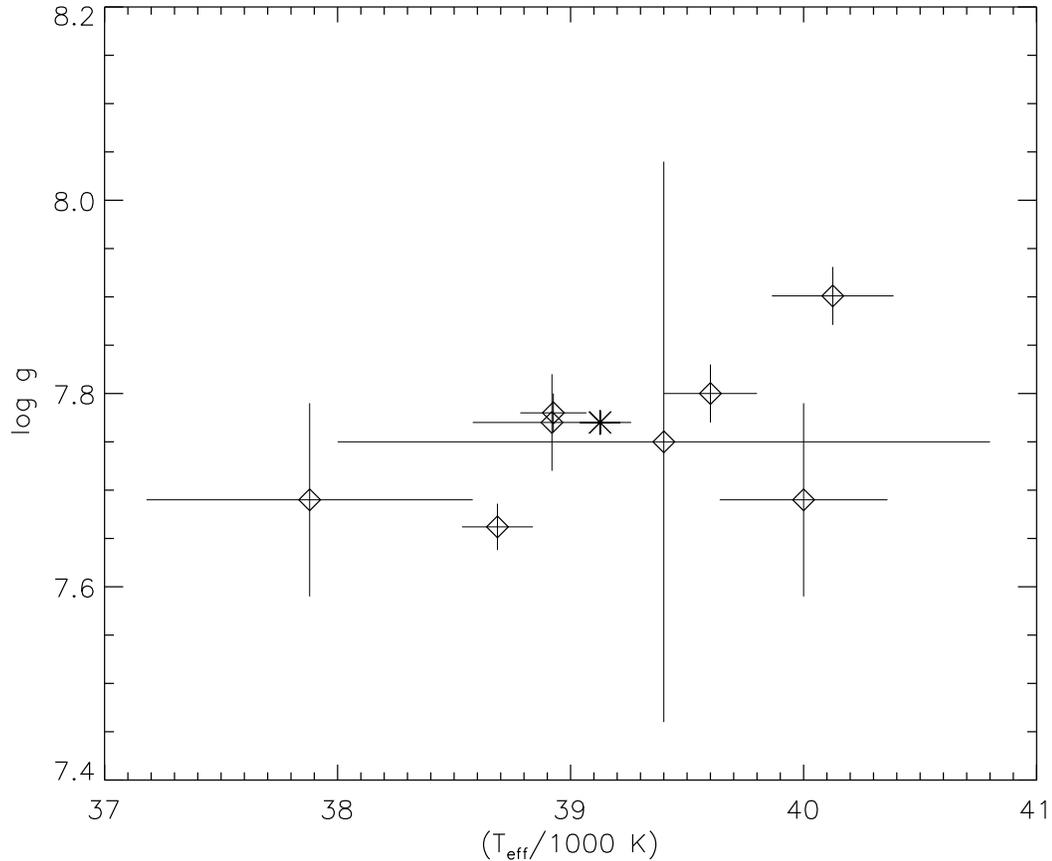}
\caption{Published independent ground-based Balmer line measurements
of \teff \ and $\log g$ for the white dwarf GD153.}
\label{gd153}
\end{figure*}

\begin{figure*}
\leavevmode\epsffile{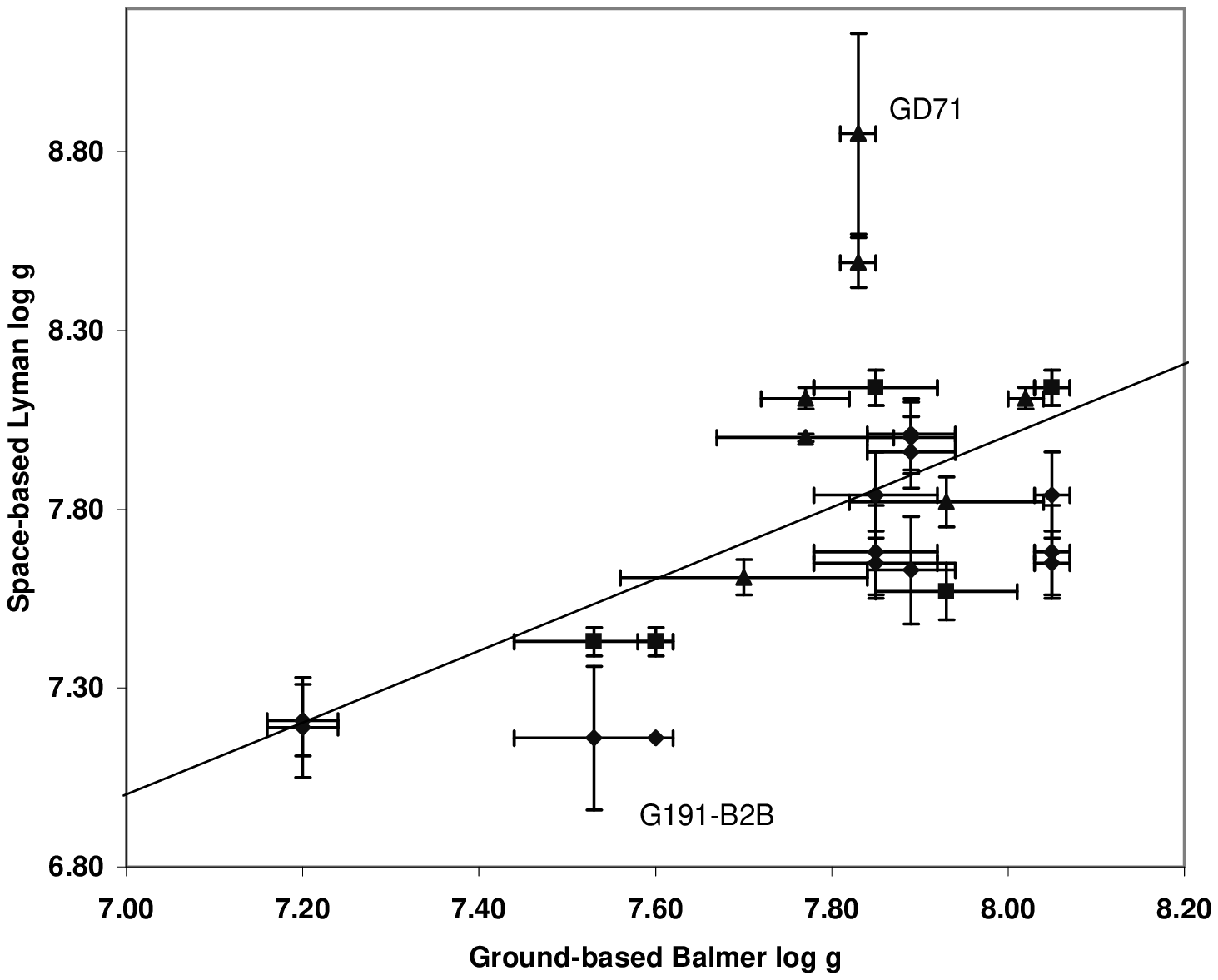}
\caption{
Scatter plot of the values of $\log g$ measured using the ground-based
Balmer (x-axis) and far- UV Lyman (y-axis) series lines. The 
diamonds are \hut \ data, the squares are the \orf \ data
and the diamonds the \fu \ data. The error bars displayed
correspond to statistical $1\sigma $ uncertainties. The light solid line
corresponds to equal Balmer and Lyman line gravities.}
\label{gbfuv2}
\end{figure*}

The information included in figure~\ref{all} does not discriminate
between the source of optical observations 
used for comparions with the far-UV  measurements. Although
the number of repeated optical observations is small, 
it is still relevant to look
for systematic differences to assess the magnitude
of systematic uncertainties in the determination of \teff. 
Figure~\ref{gbgb} compares the
Balmer line temperatures measured using ``old" spectra from our archive
(x-axis) with those of improved s/n from the more recent (``new") observations. 
It is clear that, compared to the size of the formal errors, the two sets of data,
give different results. Observations for the same stars analysed with the same
models disagree for all four stars. The differences range from 5-10\%, 
but are not all in the same direction. In fact, the best fit
straight line through the data points is y=1.01x, only 1\% \ different
from the ideal y=x. This result would suggest that the ground-based
temperature measurements can have systematic errors at the 5--10\% \
level, which dominate the statistical errors in spectra of the
signal-to-noise used here. Since \teff \ and $\log g$ are correlated,
similar systematic effects exist for the $\log g$ estimates. 
This is not a new result and Bergeron \etal \ (1992) use multiple
observations of the same star to quantify the systematic errors.
The problem is illustrated in figure~\ref{gd153}, which shows a summary
of spectroscopic observations of GD153. This suggests that, for ground-based \teff \
and $\log g$ measurements, the statistical uncertainties underestimate
the true error by a factor of 2 to 3. Nevertheless, the statistical
errors do provide a measure of the spectral s/n, the goodness of fit and the sensitivity
of the results to the line strength and should be reported. However,
it is not appropriate to take them too literally, particularly when
propagating them through calculations such as mass and radius
determinations.


Figure~\ref{all} shows values of \teff \ derived from
the various Lyman line analyses and each 
far-UV telescope is denoted by a distinctive
symbol (diamonds -
\hut , squares - \orf , triangles - \fu ).
It is possible to identify particular trends associated with each
instrument. Most of the \fu \ \teff \ values are in reasonable
agreement with the Balmer line measurements. The only dramatic
departure is PG1342+444 which, as noted earlier, has a 12000K lower
Lyman line temperature. There are too few \orf \ data points, over too
narrow a temperature range, to say much about the trends in this
instrument. Both HZ43 and G191$-$B2B are in agreement with
the Balmer measurements while the Lyman temperature of REJ0457$-$281
is 8000K too high. However, we have noted in our earlier work (Barstow
\etal \ 1998), that the separate Balmer and Lyman line fits for this
star are not, from the point of view of the statistical
errors and the goodness of fit (as determined by the F-test
described in Barstow \etal \ 1998), significantly different 
from the average fit to both data
sets simultaneously. If we a try a combined fit for
PG1342, we do not get the same result: the Lyman, Balmer and combined
fits are all significantly different using the same statistical criteria. 
As discussed earlier in this section, the statistical errors
are not a very good indication of the true uncertainty when systematic
effects are considered. If we adopt a factor 3 scaling of the
statistical errors as representative of the overall analysis process,
the level of disagreement between Lyman and Balmer analyses is within
$2\sigma $.

Considering the whole sample of observations, the \hut \ data seem to yield
Lyman line temperatures that are systematically lower than the Balmer
line values. This is most marked in GD394, but is also seen in
HZ43 and GD50. Dupuis \etal \ (2000) have already noted this in their
detailed study of GD394. Although the discrepancies observed in the
other two stars are smaller, it may not be possible to rule out a
unique systematic problem with the GD394 data. Hence, the observed
temperature effect might not be real. The lower resolution of the \hut
\ data, when compared to the other instruments pose a particular
problem, as outlined earlier, in terms of taking account of geocoronal
contamination and interstellar absorption. In addition, removing the
entire line cores to eliminate these effects does appear to have a
systematic influence on the outcome of the Lyman line analysis.

\subsection{Surface gravity measurements}

Most of the emphasis of this analysis and discussion has been on the
comparison of temperature measurements. However, in determining the
evolutionary status and important parameters such as mass and radius,
the measurement of surface gravity is of equal importance.
Figure~\ref{gbfuv2} shows the measured surface gravities corresponding
to the temperature data in figure~\ref{all}. The fractional
statistical errors on $\log g$ are larger than for the temperatures
and most error bars overlap the line of equally Balmer and Lyman line
gravity. However, there are some marked departures from this. The
surface gravity of GD71 obtained from the Lyman line analysis is
approximately 1.0dex greater than the Balmer line result, which does
not seem reasonable for such a well-studied star. On the other hand
the surface gravity measured for G191$-$B2B is much lower (by 0.37dex)
than the Balmer line result. Like the differences in \teff \ values,
these large differences in $\log g$ probably arise from systematic
effects in the data. However, there are no apparent trends for any
particular instrument and, as in the temperature measurements, no
overall systematic departures from the line of equal $\log g$.

\begin{figure*}
\leavevmode\epsffile{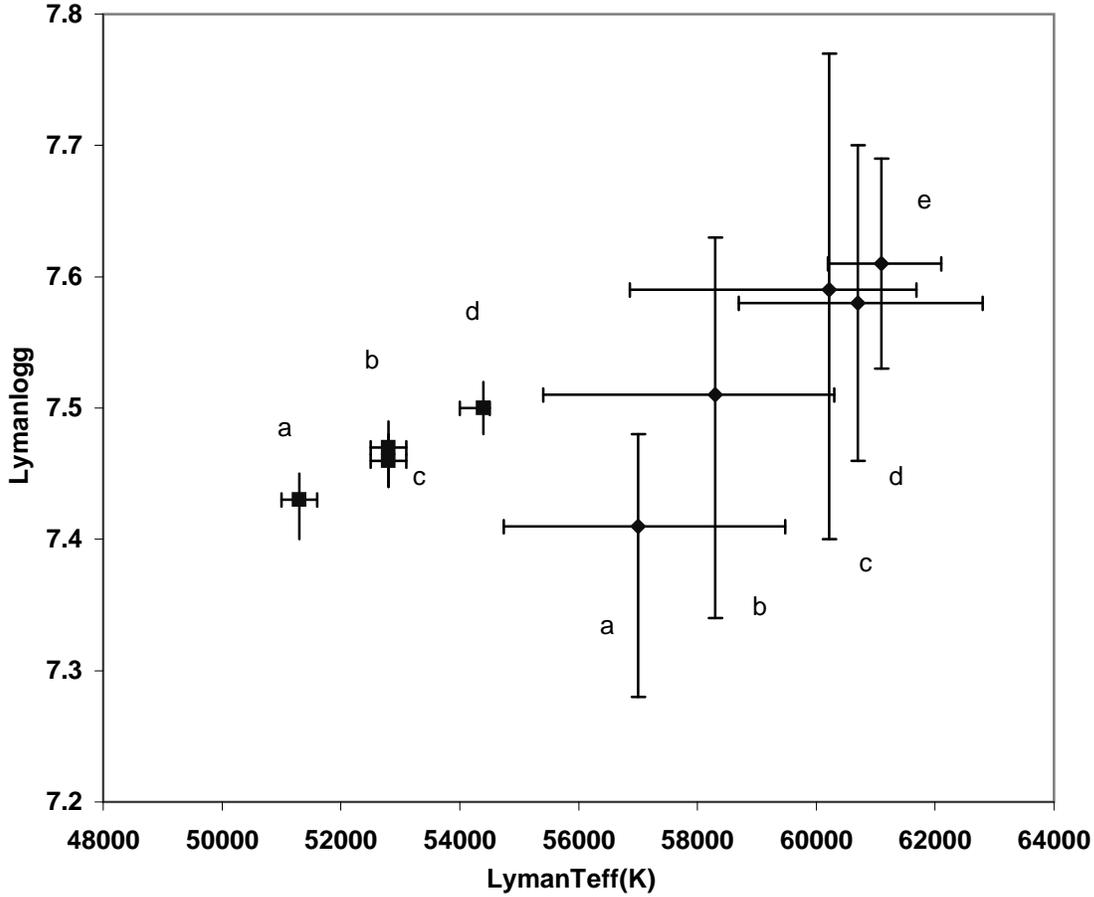}
\caption{Scatter plot of the values of \teff  and $\log g$ measured 
for the simulated \hu \ (diamonds) and \orf \ (squares) data sets.
Each data point is labelled with a letter as indicated in table~\ref{sim}.
}
\label{simf}
\end{figure*}

\subsection{Systematic effects in the analyis of \teff \ and $\log g$}

One of the main propositions of this paper is that measurements of
\teff \ and $\log g$ using either Balmer or Lyman series lines are
prone to systematic errors that are not usually well quantified. For
the Lyman lines, there is the added complication of geocoronal
emission and interstellar absorption modifying the line cores, besides
any instrumental effects such as scattered light/background
contamination. Consequently, we have simulated some of these effects
for selected \hu \ and \orf \ data sets to examine the possible
contribution to the measurement uncertainties. Fluxes were scaled to
analyse the effect of errors in effective area calibrations and
constant fluxes were added/subtracted to study the sensitivity to
accurate background subtraction/scattered light corrections. For the
\hu \ simulation, we also carried out an analysis including the
unresolved line cores. In each case, measurements were carried out
using the spectral analysis techniques described in section for the
real data. Table~\ref{sim} summarises the results, which are also
shown in figure~\ref{simf}.

\begin{table*}
\caption{Values of $\teff $ and $\log g$ measured 
for the simulated \hu \ and \orf \ data sets.
Letter labels correspond to the data points in figure~\ref{simf}.}
\begin{tabular}{lll}
& \hu \ simulation  \\
\teff (K) (err) & $\log g$  (err) & index/comment \\
57000(2400) & 7.41(0.20) & a) basic data \\
58300(2500) & 7.51(0.15) & b) data $-$ constant 5\% \ of mean flux\\
60220(2300) & 7.59(0.14) & c) data scaled by factor 1.05\\
60700(2050) & 7.58(0.12) & d) data $+$ constant 5\% \ of mean flux\\
61100(950)  & 7.61(0.08) & e) basic data, but line cores included \\ 
& \orf \  simulation\\
\teff (K) (err) & $\log g$  (err) & index/comment \\ 
51300(300) & 7.43(0.03) & a) data $-$ constant 5\% \ of mean flux\\
52800(300) & 7.47(0.03) & b) basic data\\
52800(300) & 7.46(0.02) & c) data scaled by factor 1.05\\
54400(250) & 7.50(0.02) & d) data $+$ constant 5\% \ of mean flux\\
\end{tabular}
\label{sim}
\end{table*}

It is clear that there are significant differences between the measurements
of \teff \ and $\log g$ for each of the simulated datasets. The systematic
effects that have been introduced are typically $\approx 5$\% \ of the mean
flux. These translate into observational errors $\approx 2-5$\% \ for \teff
\ and $\approx 0.5-1.0$\% \ for $\log g$. While the precision of the \hu \
measurements is similar to the level of the systematic errors, the systematic
effects clearly dominate the statistical uncertainties in the \orf \
analyses. It is interesting to note that the total spread of temperature
measurements ($\approx 10000$K) for the simulated data set is as large
as some of the perceived discrepancies in the real observations.

\section{Conclusion}

We have measured the effective temperatures and surface gravities for
a sample of hot DA white dwarfs  using the Lyman line data available
from the \hut , \orf  \ and \fu \ far-UV space missions, comparing the
results with those from the standard  Balmer line technique. In
general, there is good overall agreement  between the two methods.
At the level of the pure statistical errors, differences are found 
between Lyman and Balmer measurements for a number of stars. However, 
across the sample, the
discrepancies are not always in the same direction. This is not what
would be expected if the problems arose from limitations of the
stellar atmosphere calculations and, in particular, the treatment
of Lyman and Balmer line
broadening. Hence, while the description of opacity in the region
of the Lyman lines is still uncertain, because of uncertainties
associated with the occupation probability formalism, the
description of the line broadening of the isolated, lower, Lyman
lines seems to be in good shape.
It is not  necessary to make semi-arbitrary
adjustments to the input physics, as carried out by
Finley \etal \ (1997b) in their analysis of the \hut \ 
data, to force agreement between the Balmer
and Lyman line results. Even so, there may well be residual second
order effects worthy of examination.  The best fit line in
figure~\ref{all} (excluding problematic results noted earlier) hints
at a few percent systematic difference between Balmer and Lyman values
of \teff . However, this is at a similar level to the possible
systematic data reduction problems discussed above, which must be
eliminated before we can consider this further.

It is well known that optical Balmer line measurements can be
sensitive to systematic effects in the data  reduction process, in
particular the extinction correction that must be applied to take
account of  atmospheric absorption as well as the absolute flux
calibration. Although the former problem is not  present in the far-UV
spectra it is apparent that this method of temperature and gravity
measurement is  sensitive to a number of systematic effects that can
compromise these results. For example, with the lower  spectral
resolution \hut \ data, the geocoronal and interstellar contamination
of the line cores cannot be  dealt with very easily, leading to a
systematic shift in measured values of \teff . In a simulation,
inclusion of the line core regions in the analysis yielded erroneously
high values of \teff \ and $\log g$. In other cases, the accuracy of
subtraction of instrument background or scattered light may be
important. Simulations of over- or under-subtracted background
components at the level of 5\% \ of the mean flux give shifts of
similar magnitude in the measured value of \teff .  With high
signal-to-noise data available from a variety of instrumentation, it
is clear that our analyses are  no longer limited by the statistical
errors but by the systematic ones arising in the process of data
acquisition and reduction. Consequently, we conclude that, if these
effects can be adequately controlled,  the Lyman line technique gives
measurement that are consistent with those obtained from the Balmer lines,
when allowance is made for realistic observational uncertainties.


\section*{Acknowledgements}

The work reported in the papers was based on observations made with the \hut,
\orf \  and \fu \ 
observatories. We would like to thank Jeff Kruk and Mark Hurwitz for supplying
the \hut \ and \orf \ spectra respectively.
MAB and SAG were supported by PPARC, UK. JBH wishes to acknowledge support for 
this work from NASA grants NAG5-9181 and NAG5-8995. FM was supported by a Nuffield Foundation 
Student Bursary. We would like to thank the referee, Pierre Bergeron,
for his very careful critique of this paper and his valuable suggestions
for its improvement.

\end{document}